\documentclass[12pt,onecolumn]{article}
\usepackage[utf8]{inputenc}
\usepackage{geometry}
\usepackage{times} 
\usepackage{graphicx} 
\usepackage{subfig}
\usepackage{svg}
\usepackage[bookmarks=false]{hyperref}
\usepackage{cite}
\usepackage{amsmath} 
\usepackage{lipsum}
\usepackage{caption}
\usepackage{fancyhdr} 
\usepackage{titlesec} 
\titleformat{\subsection}
  {\normalfont\bfseries\itshape}{}{0em}{}
\date{}
\title{\textbf{SDR-Based Metal Classification using Spectrogram Images from Micro-Doppler Signatures}}

\author{
    Salman Liaquat, Faran Awais Butt$^*$, Faryal Aurooj Nasir, Ijaz Haider Naqvi,\\ 
    Nor Muzlifah Mahyuddin$^*$, Ali Hussein Muqaibel, and Saleh Alawsh
    \thanks{Corresponding authors: Faran Awais Butt (faranawais.butt@kfupm.edu.sa) and Nor Muzlifah Mahyuddin (eemnmuzlifah@usm.my)}
}

\begin{document}

\maketitle
\thispagestyle{fancy}

Metallic materials such as brass, copper, and aluminum are used in numerous applications, including industrial manufacturing. The vibration characteristics of these objects are unique and can be used to identify these objects from a distance. This research presents a methodology for detecting and classifying these metallic objects using the vibration dynamics induced by their micro-Doppler signatures. The proposed approach utilizes image processing techniques to extract pivotal features from spectrograms. These spectrograms originate from micro-Doppler signatures of data collected during controlled laboratory experiments where signals were transmitted towards vibrating metal sheets, and the ensuing reflections were recorded using a software-defined radio (SDR). The spectrogram data was augmented using geometric transformation to train a convolutional neural network (CNN) based machine learning model for object classification. The results indicate that the proposed CNN model achieved an accuracy of more than 95\% in classifying metals into brass, copper, and aluminum. This research could be used to understand the foundations of classifying spectrogram images using micro-Doppler signatures for its applications towards enhancing the sensing capabilities in industrial and defense applications.  

\section*{Introduction}

Metallic objects are used in several industrial applications due to their excellent mechanical properties, including high strength, ductility, and toughness. However, the mechanical behavior of metals can be affected by external factors, such as vibration, which can cause changes in their microstructure and properties. Vibration-induced changes in metals can occur for several reasons, including dynamic loading, fatigue, and wear. When a metal is subjected to vibration, its microstructure undergoes deformation, resulting in changes in grain size, dislocation density, and residual stress\cite{huang2021microscopic}. These changes can affect the mechanical properties of the metal, leading to decreased strength, ductility, and toughness.
  
Micro-Doppler research has been widely used in biomedical and disaster management applications, including detecting vital/life signs. Due to their stability and robustness, micro-Doppler characteristics can be used to extract unique signatures of a target. Micro-Doppler signatures have garnered significant attention in scholarly research because of their potential to identify targets that exhibit micromotions \cite{hanif2022micro}. 
Micro-Doppler signatures have been used for numerous purposes, from enhancing personnel recognition and gait classification accuracy using multistatic radar micro-Doppler signatures with deep convolutional neural networks (CNNs)  to discriminating between unarmed and potentially armed personnel \cite{8307105}. The efficient extraction of micro-Doppler features from Doppler radar data has been used in classifying different human activities. Devices such as radar systems, laser vibrometers, and microwave Doppler radar have been effectively employed to measure micro-Doppler effects. For instance, continuous wave radars, frequency modulated continuous wave (FMCW) radars, and ultra-wideband radars have been utilized in diverse applications ranging from personnel recognition to vibration analysis in structures. These systems are capable of capturing detailed micro-Doppler signatures, which can then be analyzed to understand the dynamic behavior of different materials and objects.

The focus of the proposed technique is on classifying metals based on their vibration signatures, using the strengths of micro-Doppler radar in combination with machine learning techniques. This approach can be particularly suitable for applications in recycling, manufacturing quality control, and defense settings, where the ability to remotely classify materials is of paramount importance, making it a valuable addition to the existing measurement techniques in the instrumentation and measurement (IM) field.

\subsection{Doppler and Micro-Doppler Effects in Radar}

In radar target detection, the Doppler effect primarily ascertains the radial velocity of a target, whereas the micro-Doppler effect offers insights into the target's characteristics. Both effects manifest in the radar signal received after striking the target. While the Doppler frequency sheds light on the motion of the target, the micro-Doppler unveils the detailed dynamics of its constituent parts. The Doppler effect refers to the change in frequency of a wave with respect to the radar receiver. When a target, such as an aircraft or car, moves relative to the radar, the frequency of the reflected wave differs from that of the transmitted wave from pulse to pulse. The plot of pulses over time is known as slow-time in radar and the frequency variation due to the moving object is termed the Doppler shift, $f_d$, or Doppler frequency and for a target approaching the radar with a velocity $v_r << c$, it is given by \cite{muqtadir2021health},

\begin{equation} 
f_d =\frac{2v_r}{\lambda},
\label{Doppler shift} 
\end{equation}

\noindent where $v_r$ is the target radial velocity and $\lambda$ is the wavelength of the transmitted signal. The Doppler shift depends on $\lambda$ and changes with a change in the radar's operating frequency. The same radial velocity corresponds to a higher Doppler shift for a higher frequency of operation. The micro-Doppler effect describes the frequency modulation that results from the rotational or vibrational movements of a target or its components. It represents the micro-Doppler effects superimposed on the primary Doppler effect. For example, the rotating blades of a helicopter or the legs of a walking individual introduce micro-Doppler shifts in the radar's returned signal. If we consider a point on a rotating object, its radial velocity due to rotation is,

\begin{equation} 
v_r = \omega r,
\label{microDoppler} 
\end{equation}

\noindent where $\omega$ is the angular speed of rotation (radians per second) and $r$ is the distance of the point from the axis of rotation. The micro-Doppler frequency $f_{md}$ due to this rotation is given by,

\begin{equation} 
f_{md} =\frac{2\omega r}{\lambda}.
\label{microDoppler frequency} 
\end{equation}

Micro-Doppler radar plays a significant role in the field of IM due to its ability to capture fine motion details of objects. This technology is particularly valuable in applications such as structural health monitoring, vibration analysis, and material characterization. By analyzing the micro-Doppler signatures, one can extract critical information about an object's dynamic behavior, which is essential for accurate measurement and analysis in various industrial and scientific contexts. Since early applications of radar micro-Doppler signatures for human gait studies, tracking bedridden patient positions and person identification based on sit-to-stand and stand-to-sit movements using micro-Doppler radar measurements and a CNN \cite{9004533}, they are being used in advanced applications, including recognition of targets in space (for eg. missile warheads), recognition of targets in the air (like flying birds), recognition of ground moving targets (vehicles, for instance), recognition of targets underground (such as vital sign detection), and recognition of targets behind a wall. Microwave-based radar sensors are also being used for healthcare and security applications \cite{muqtadir2021health}. Deep neural networks (DNNs) have emerged as a significant topic in the classification of radio frequency signals and the analysis of remote sensing data \cite{10004894}. The use of CNN with micro-Doppler signatures has been used to classify real-time fall recognition and flying objects classification \cite{10401166}. Data augmentation techniques offer a highly effective strategy for mitigating model overfitting throughout the network training phase in all the situations it has been applied.
The spectrogram representations can take advantage of the micro-Doppler signatures, serving as distinctive features for classification. For instance, the radar micro-Doppler signature-based approach can be used for non-cooperative target identification, recognition, and classification \cite{khalifa2022comprehensive}. 

While studying the micro-Doppler signatures, it is crucial to consider the surface vibrations of objects. By incorporating vibration information, our research aims to provide a more comprehensive analysis of the micro-Doppler signatures. This approach enables us to capture detailed vibration patterns that are unique to each type of metal, thereby improving the accuracy and reliability of our classification method. This method is particularly beneficial when a preliminary assessment of the metal type is satisfactory, particularly in the early recycling phases or during initial sorting procedures on production lines. Quick and accurate identification of metals can streamline the sorting process, reduce operational costs, and improve overall efficiency. In this research, we investigate the micro-Doppler signatures of vibrations induced by different materials used in engineering and high-voltage equipment.
An electromagnetic wave is transmitted from a software-defined radio (SDR) transmitter connected to a directional antenna. The transmitted signal gets reflected and scattered and sometimes penetrates the object. The received signal at the radar is then processed to investigate the variations of the micro-Doppler signatures of the target, which helps to classify the targets based on their unique micro-Doppler signatures. Targets exhibit different micro-Doppler signatures based on the materials they are composed of. This presents a potential application for object identification and classification.\\

\subsection*{Applications and Future Prospects}
Micro-Doppler radar has been utilized in various innovative applications across different fields. In healthcare, it is employed for monitoring vital signs and movements of patients, providing a non-invasive way to track health status. Structural health monitoring also benefits from this technology through the assessment of the integrity of buildings and bridges via vibration analysis, which aids in the early detection of structural issues.
In the manufacturing industry, quality control is critical to ensure that materials and components meet specified standards. The proposed micro-Doppler radar technique can be employed in a production line setting to classify metals. By analyzing the unique vibration signatures of different metals, the system can quickly identify and sort materials. For example, during the production of metal components, this method can be used to ensure that only metals with the correct properties proceed to the next stage of production. This application not only demonstrates the practical utility of the methodology but also highlights its potential to improve operational efficiency and product quality in industrial settings.
In the security domain, micro-Doppler radar is particularly useful for detecting and classifying drones and other aerial threats. By analyzing the unique micro-Doppler signatures of different types of drones, the radar system can accurately identify potential threats from a distance, enhancing surveillance capabilities and enabling timely responses to security breaches.

The materials chosen for this study—brass, copper, and aluminum—were selected due to their widespread industrial usage and their unique mechanical and physical properties. Brass, copper, and aluminum are commonly used in a variety of applications, including construction, manufacturing, and electronics. These metals exhibit distinct micro-Doppler signatures due to differences in density, elasticity, and surface characteristics. By selecting these materials, the study aims to demonstrate the effectiveness of micro-Doppler radar in distinguishing between metals with varying properties, which is crucial for applications in quality control, material identification, and industrial automation.

\section*{System Model}

\begin{figure*}[t!]	
 	\centering
 	\includegraphics[width=\linewidth]{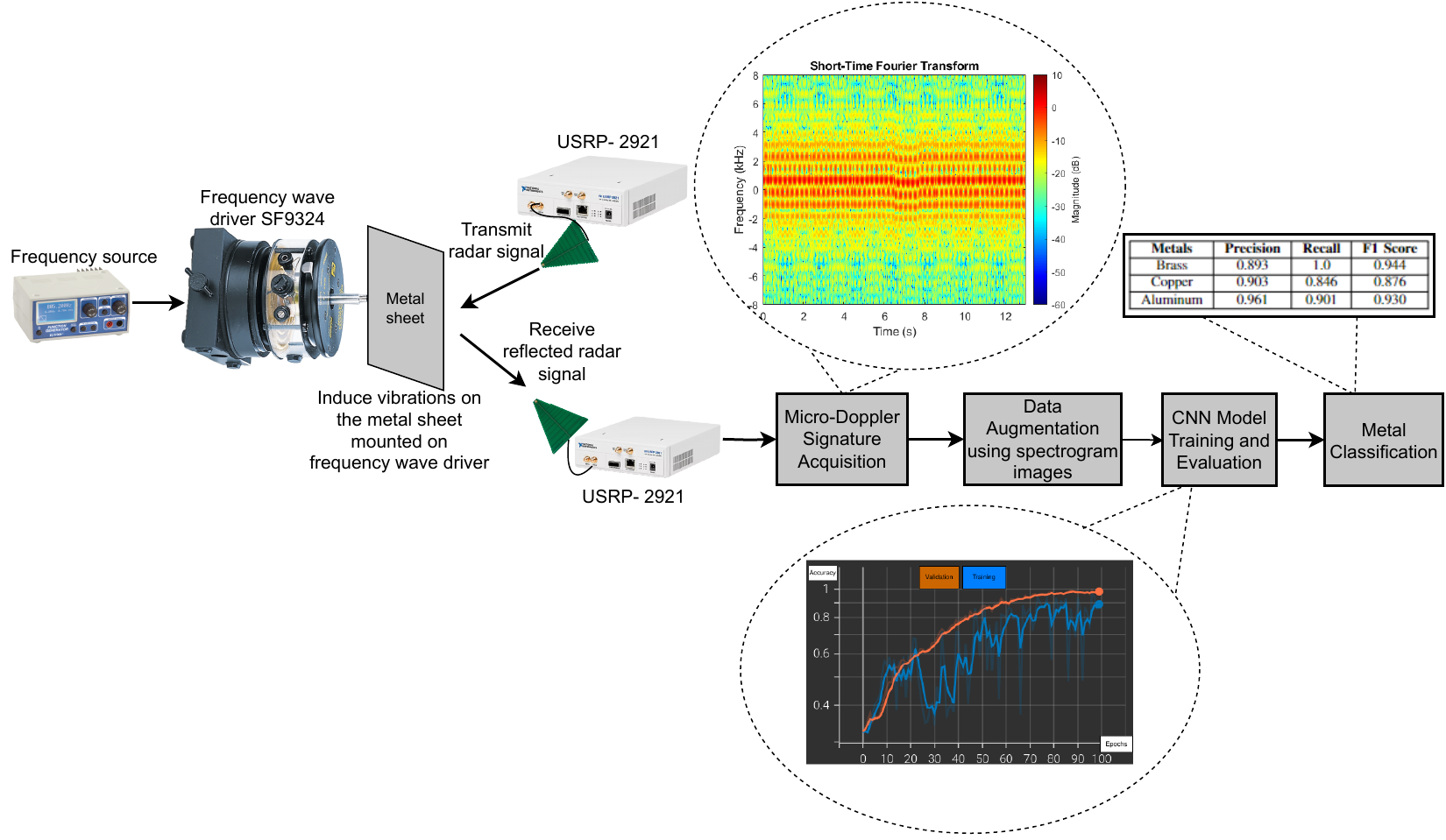}	
 	\caption{Functional block diagram illustrating the classification of micro-Doppler signatures.}
	\label{Graphical.pdf}	   
    \end{figure*}

A variable frequency wave driver was used to induce vibrations in sheets of these metals, and two National Instruments (NI) Universal Software Radio Peripherals (USRPs) 2921 were employed to transmit and receive the signals. The acquired signal data was then processed, which enabled the extraction of micro-Doppler signatures. The signatures in the form of spectrogram images are then resized, followed by a data augmentation process because of a limited dataset. A CNN-based model was then trained for metallic object classification, and the results have been promising, allowing us to propose the use of machine learning in metallic object classification. The sequential breakdown of the proposed methodology for metal object classification via radar-based micro-Doppler image analysis using CNN is depicted in Fig. \ref{Graphical.pdf}. After acquiring the reflected signals from the target, we employ MATLAB for signal processing. We initiate a time-frequency analysis of the incoming signal, calculating its short time Fourier transform (STFT). The STFT is applied to the radar signals to convert them from the time domain into the frequency domain. This transformation is crucial for visualizing the micro-Doppler effects, as it allows the detection of subtle frequency shifts caused by the micro-motions of the vibrating metal objects. The resulting spectrograms, which display these frequency variations over time, reveal the unique micro-Doppler signatures of the materials under study. These signatures are essential for accurately classifying the materials based on their vibration characteristics. This transform captures distinct vibrational signatures inherent to various materials, enabling object classification using micro-Doppler signatures.  The spectrogram data was augmented through the application of geometric transformations, which were utilized to train a CNN-based machine learning model for the purpose of object classification.

\section*{Experimental Setup and Proposed Methodology}

\begin{figure}[t!]	
	\centering
	\includegraphics[width= \linewidth]{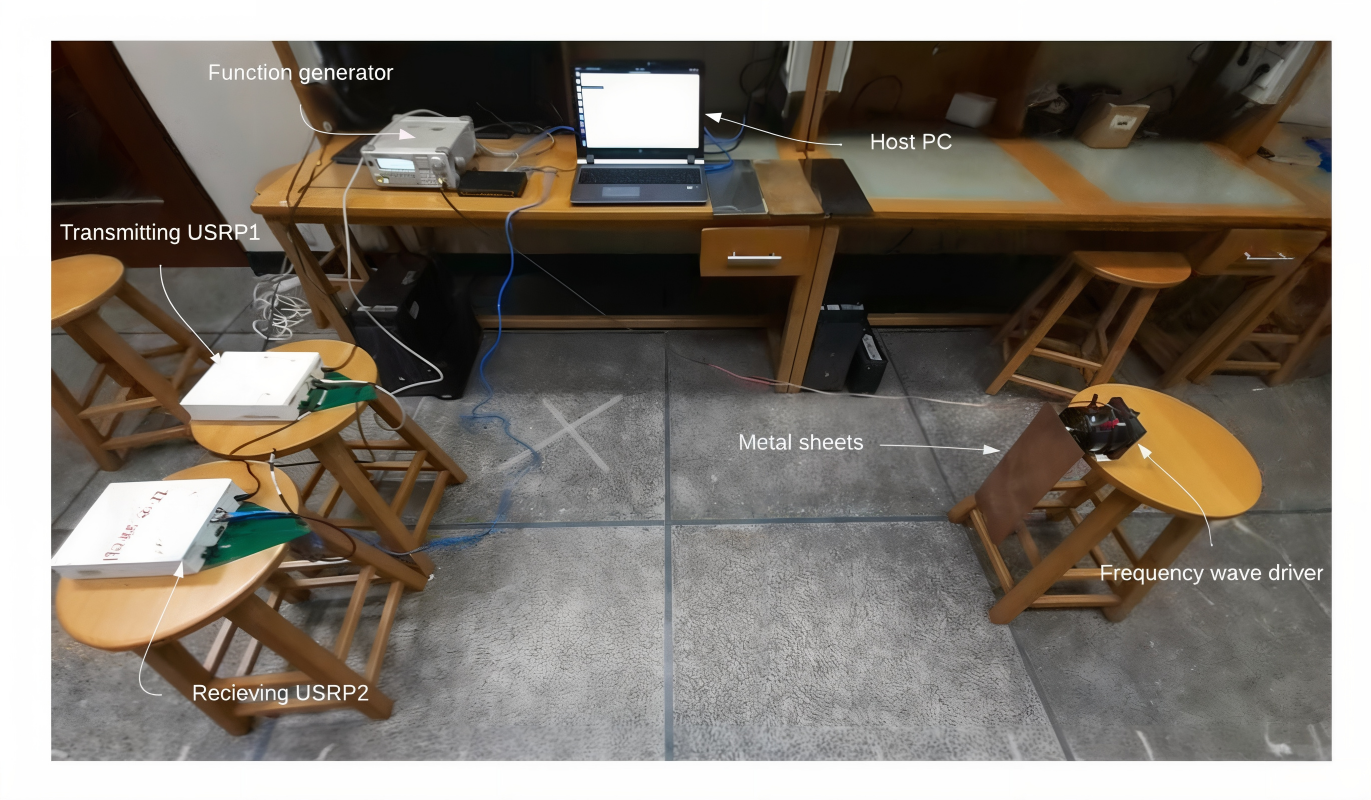}	
	\caption{Experimental setup for acquiring micro-Doppler signatures.}
	\label{su}	
\end{figure}

A frequency source with a variable frequency wave driver, SF9324 by PASCO Scientific, was used to induce the vibrations in metallic sheets. The thickness of all used sheets was set to $0.37~mm$, with dimensions of $36~cm \times 14~cm$. Larger or thicker sheets may have lower natural frequencies and could respond differently to the induced vibrations, impacting the effectiveness of the frequency used in the experiment. Materials with larger dimensions or those that are more reflective can have a higher radar cross section (RCS), improving signal detection but possibly also introducing more signal reflections and multipath effects. Fig. \ref{su} shows the overall setup where two USRP boards (NI 2921) were used as a transmitter and a receiver at 2.45 GHz. The figure provides a visual representation of the experimental setup, showing how the metal sheet is mounted to the driver and other critical setup parameters, such as the height and distance between components. The setup height was approximately $1~m$ and the distance between the USRPs and the metallic sheets was set to $1.5~m$ in order to detect the vibrations. The distance between the transmitter and the receiver antenna is $0.3~m$ to ensure the receiver captures only the reflected signal. The sheet under examination was mechanically coupled with the driver, ensuring direct contact. This setup ensures optimal energy transfer to induce the desired vibrations. Proper contact is crucial: insufficient contact may lead to inadequate energy transfer, while overly tight contact could cause excessive localized stress or potential damage to the sheet. Direct contact is essential for achieving accurate vibration induction and reliable measurement results. The distance between the USRPs also impacts the signal path length between the transmitter and receiver, affecting signal strength and quality. Moreover, directional log periodic antennas were employed at both the transmitter and receiver to ensure accurate transmission and reception of signals within our experimental framework.

We chose to generate a sinusoidal signal of 200 Hz to simulate the sinusoidal vibrations often exhibited by metallic objects. This frequency was selected based on its relevance to real-world scenarios and its ability to effectively induce vibrations for our experimental setup.
The chosen specifications were selected based on the availability of materials and the need to demonstrate scalability, offering a practical and replicable framework for future studies and real-world applications. This deliberate choice allows for robust experimentation while ensuring that the findings can be easily extrapolated and applied across a range of scenarios, thereby enhancing the credibility and generalizability of the research outcomes.
Upon examination of radar return signals within the frequency domain using a spectrogram, it can be observed that the micro-Doppler effect manifests as sidebands surrounding the primary Doppler frequency. This primary frequency represents the overarching motion of a given object, such as a metal sheet. In situations where this sheet remains stationary, i.e., barring any induced vibrations, the primary Doppler frequency would ideally align around zero frequency, essentially coinciding with the carrier frequency. Due to the vibrations of the sheet, micro-Doppler effects present themselves as symmetric sidebands. These are distanced from the central frequency by a magnitude corresponding to the vibration frequency.

\subsection{Proposed Methodology}
Radar signals that were reflected off vibrating metal sheets were acquired and transformed into images called spectrograms, which show how the signal changes over time. These spectrogram images were then used to train a CNN, a type of machine learning model. The CNN learns to recognize patterns in these images that are unique to each type of metal. Once trained, the CNN can analyze a new spectrogram image and accurately identify the type of metal it represents. This approach allows advanced technology to classify metals quickly and accurately. \\ 
The STFT images for the vibrating metals are transformed to grayscale to simplify the analysis process. The image's contrast is then enhanced, highlighting its features for better clarity and distinction. Initially, the amplitude was mapped to a color scheme, but transforming it to grayscale helped streamline the processing steps without losing critical information. Grayscale images reduce computational complexity while preserving essential features for classification. This approach ensures that the key characteristics of the micro-Doppler signatures are maintained, facilitating accurate and efficient analysis by the CNN. A pivotal step in our analysis involves the identification of peaks within the image. These peaks correspond to the dominant frequencies in the spectrogram, revealing salient structures and patterns inherent to the dataset. Moreover, by examining the intensity distribution of these peaks, insights are derived regarding the strength of reflections from various metals. The dataset under investigation comprises images representing three metal sheet types: brass, copper, and aluminum.
The spectrogram is a high resolution image which is needed to be resized for computational purposes. The specific choice to resize images is often a balance between preserving sufficient detail for analysis and maintaining computational efficiency. For computational efficiency and uniformity, in our case, each high resolution image (\(3500\times 2625\) pixels) was resized to a resolution of \(256 \times 256\) pixels using bilinear interpolation and subsequently transformed into grayscale. For feature extraction, we employed the histogram-of-oriented-gradients descriptor, recognized for its efficacy in capturing the shape and structural nuances of objects within an image. The dataset deployed for model training comprised unmodified original images. Given the constraints of our dataset's size, to enrich the dataset and enhance the model's capacity for generalization, a series of image augmentation techniques were implemented.
Augmenting spectrograms, while not straightforward, has been successfully implemented, as demonstrated in studies such as \cite{nanni2020data}. This collection of original and systematically augmented images was then employed in training the model to observe its performance on unseen data.

The complete workflow for training and evaluating a CNN model on spectrogram images is implemented in Python.  We took eight spectrograms each for three classes of metals. We augmented them by applying geometric transformation. 
The augmentation results in the generation of ten additional images for each original spectrogram of varying resolutions (\(64 \times 64\), \(96 \times 96\), \(128 \times 128\) and \(256 \times 256\)), leading to 88 images per class after augmentation. Therefore, the total number of augmented images amounts to 264. A CNN model is defined using TensorFlow's Keras API with various layers like convolutional, batch normalization, max pooling, dropout, and dense layers. The trained model is evaluated on the testing data to measure its performance in terms of accuracy. Precision, recall, and F1-score metrics are calculated for each class based on the model predictions and true labels. The dataset of 264 samples is randomly divided into 210 samples (around 80\%) for the training and 27 samples each (around 10\% each) for validation and testing.

\subsection{Performance Evaluation Metrics}
To quantitatively analyze the performance of detection models, performance metrics such as precision, recall, and F1 score parameters are used. Precision is defined as \cite{9400955},
\begin{equation} 
\text{Precision} = \frac{TP}{TP + FP},
\label{Precision} 
\end{equation}

\noindent where \(TP\) represents the number of true positives, i.e., the instances correctly predicted as positive by the model, and \(FP\) represents the number of false positives, i.e., the instances incorrectly predicted as positive by the model.
Recall is given by,
\begin{equation} 
\text{Recall} = \frac{TP}{TP + FN},
\label{Recall} 
\end{equation}

\noindent where \(FN\) denotes the number of false negatives, i.e., the instances where the model incorrectly predicted positive instances as negative.
The F1 score, which is the harmonic mean of precision and recall, is calculated as,
\begin{equation}
\text{F1 Score} = \frac{2 \times \text{Precision} \times \text{Recall}}{\text{Precision} + \text{Recall}}.
\label{F1 Score}
\end{equation}

\noindent These metrics collectively provide a comprehensive assessment of the model's performance.

\section*{Results and Discussion}

\begin{figure}[t!]
\centering
  \subfloat[Brass]{\label{fig:bs} \includegraphics[width=0.32\linewidth]{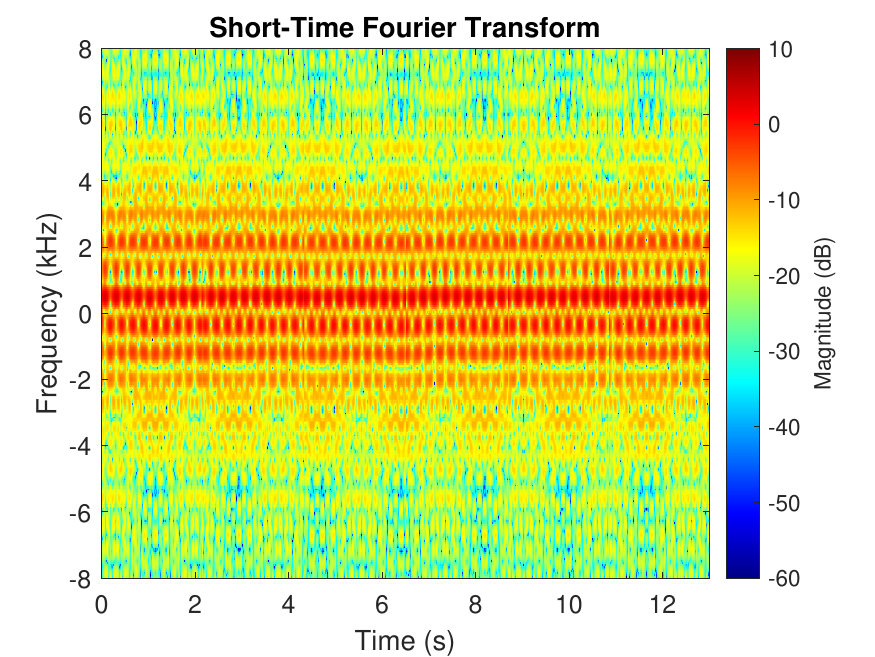}}
  \subfloat[Copper]{\label{fig:cs} \includegraphics[width=0.32\linewidth]{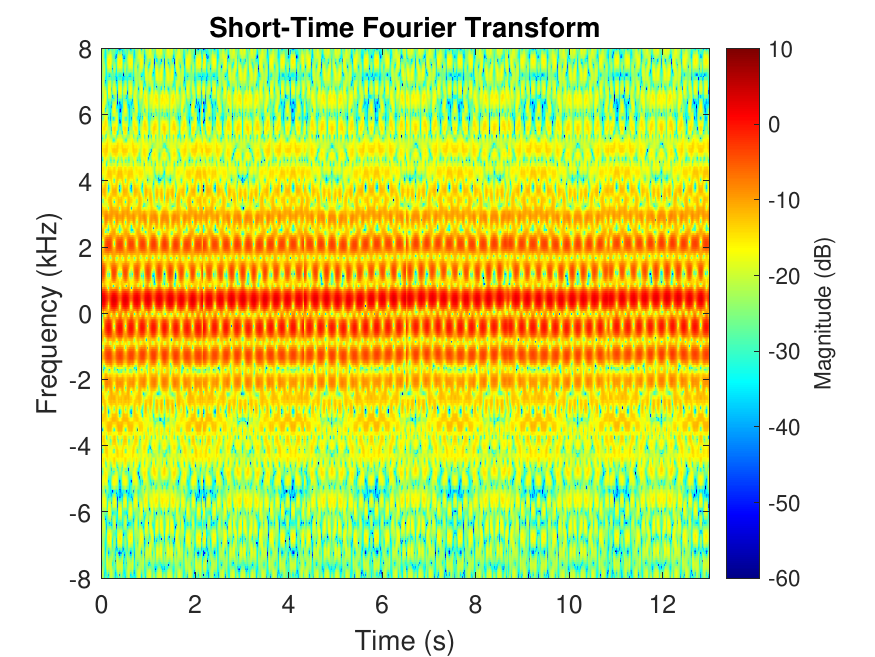}}
  \subfloat[Aluminum]{\label{fig:as} \includegraphics[width=0.32\linewidth]{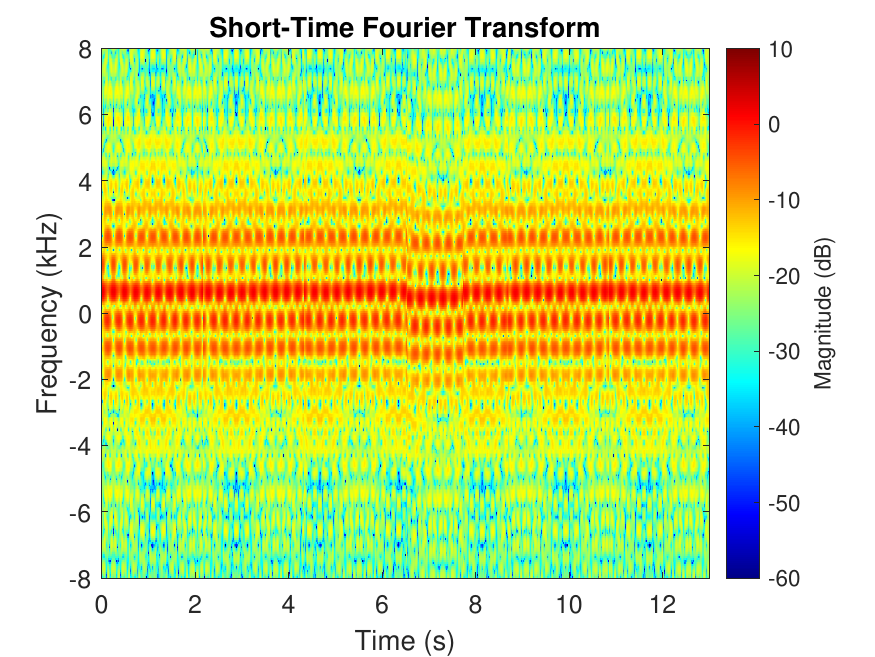}}
\caption{Sine wave spectrograms of different vibrating sheets: (a) brass, (b) copper, and (c) aluminum.}
\label{fig:vs}
\end{figure}
 
Examples of the spectrograms of vibrating brass, copper, and aluminum sheets are shown in Fig. \ref{fig:vs}. The intensity distribution in the spectrograms shows strong reflections at certain frequencies. The findings suggest that stronger reflections are observable at specific frequency components, offering potential benefits for enhanced material characterization in radar-based sensing applications. Despite having the same vibrating stimulus, the frequency of the received signal is different for all materials. Brass shows distinct frequency peaks at lower ranges, copper exhibits a broader frequency spread, and aluminum has sharper, higher frequency components. The CNN can detect and classify subtle differences in the spectrograms that may not be immediately apparent to the human eye, thereby significantly improving the accuracy of material classification. These differences are due to the unique vibration characteristics of each metal, which are captured in their micro-Doppler signatures. This distinction might not be immediately apparent to unaided observation, but becomes perceivable after applying CNN-based processing techniques, highlighting the value of computational methods in enhancing signal analysis.
\begin{table}[t]
\caption{Classification metrics for different metals.}
    \centering
    \begin{tabular}{|c|c|c|c|}
        \hline
        \textbf{Metals}   & \textbf{Precision} & \textbf{Recall} & \textbf{F1 Score} \\ 
        \hline        
        Brass    & 0.893 & 1.0 & 0.944 \\
        \hline
        Copper   & 0.903 & 0.846 & 0.874 \\
        \hline
        Aluminum & 0.961 & 0.901 & 0.930 \\
        \hline
    \end{tabular}
    \label{tab:classification_metrics}
    \vspace{-0.2cm}
\end{table}

\begin{figure}[t!]
\centering
  \subfloat[Training and Validation Accuracy]{\label{fig:accuracy} \includegraphics[width=0.48\linewidth]{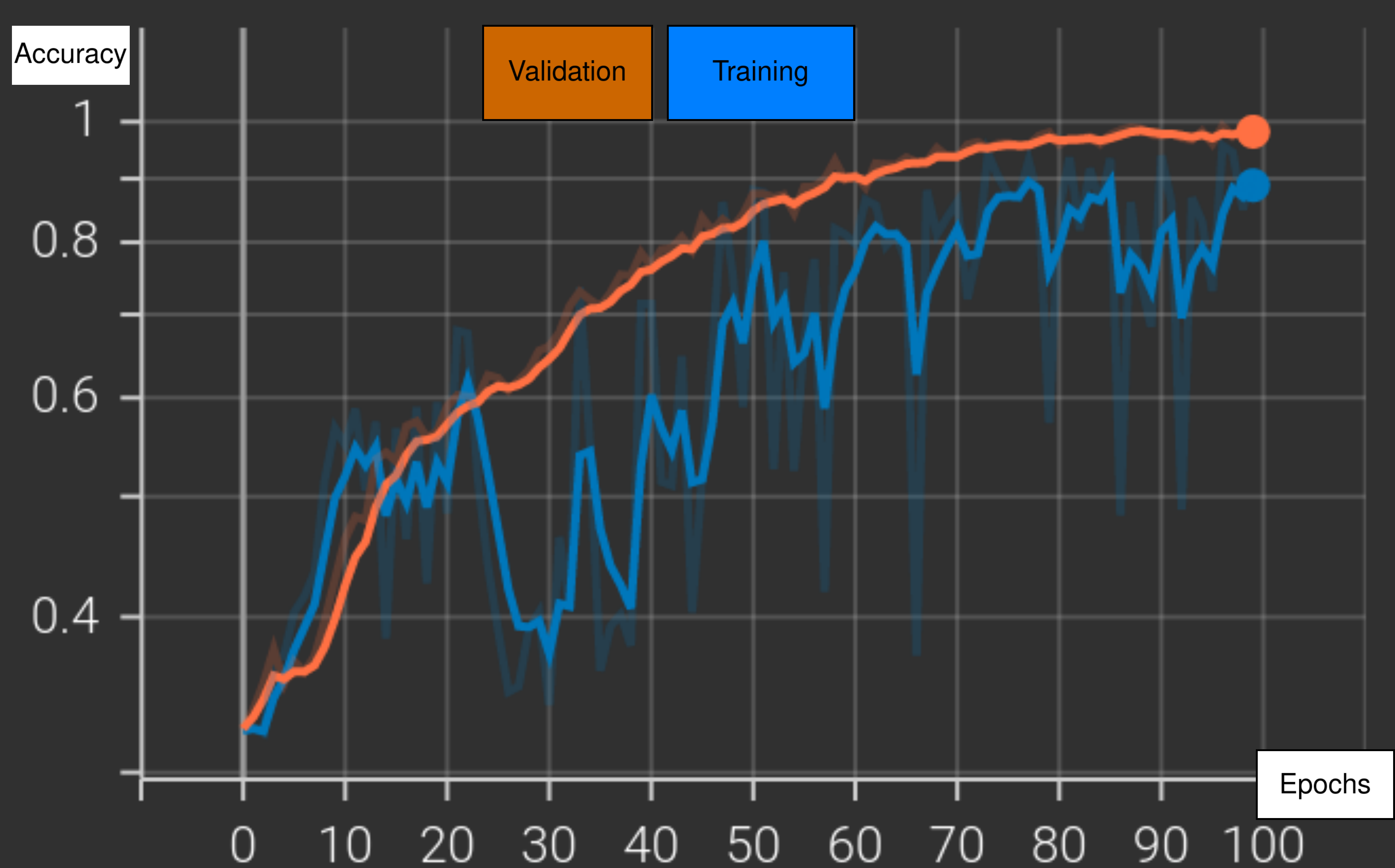}}
  \subfloat[Training and Validation Loss]{\label{fig:loss} \includegraphics[width=0.48\linewidth]{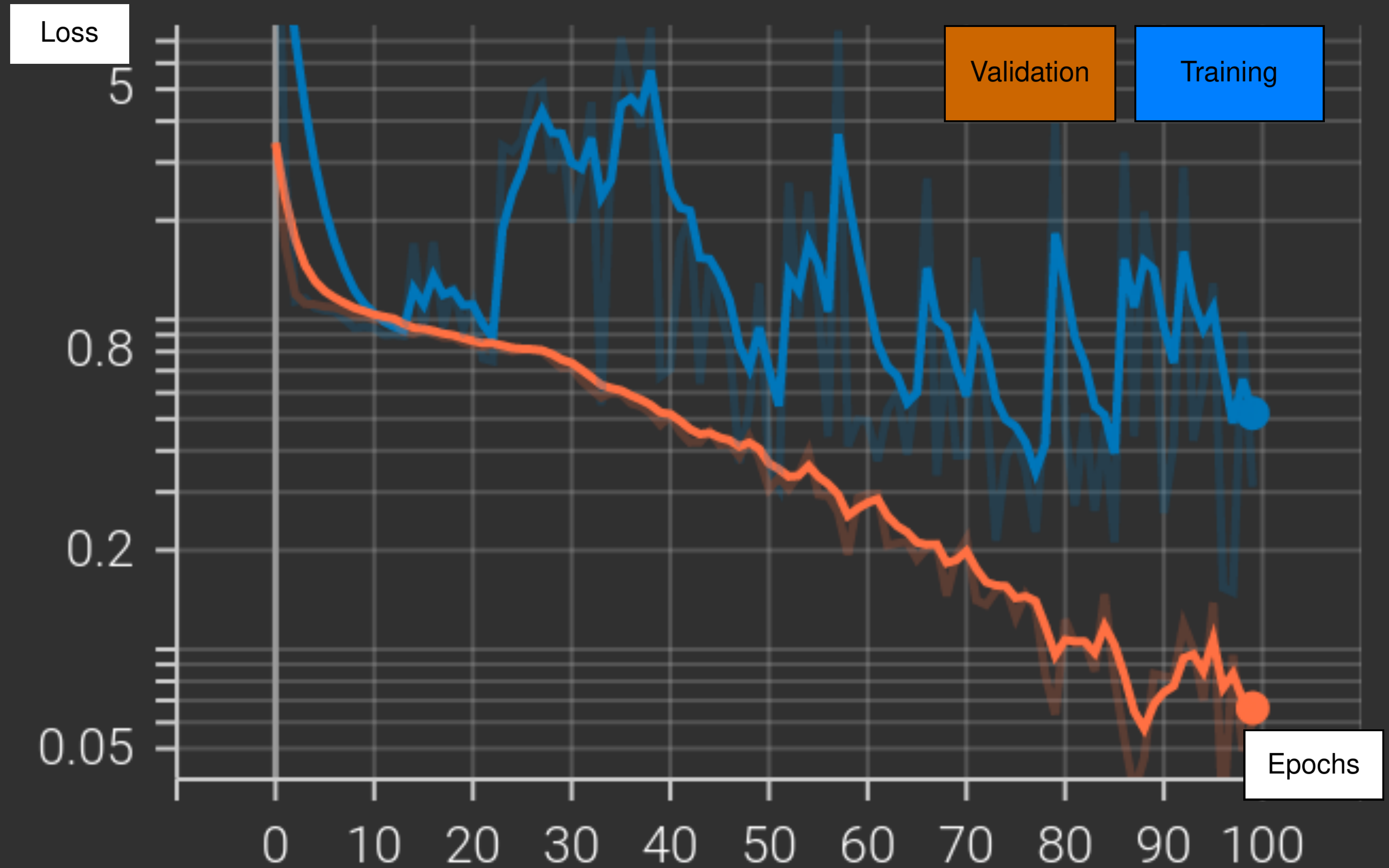}}  
\caption{Proposed CNN model's (a) Training and Validation Accuracy (b) Training and Validation Loss.}
\label{zs}
\end{figure}

In our case, the CNN model was trained on the augmented dataset and achieved a validation accuracy of more than 95\% on the test data. The classification metrics for the model and average classification metrics are presented in Table  \ref{tab:classification_metrics}. The developed approach successfully classified previously unseen data, achieving an accuracy of more than 95\% as seen in Fig. \ref{fig:accuracy}. The training and validation loss of the model is shown in Fig. \ref{fig:loss}. The accuracy and loss graphs have been plotted using TensorBoard.

\section*{Conclusion}

This study uses a CNN-based model to demonstrate the effectiveness of sensing vibrating dynamics and radar-based micro-Doppler image analysis to classify metallic objects like brass, copper, and aluminum. Our methodology utilizes dynamic platform vibrations and advanced image processing techniques to extract crucial features from spectrograms, which are then used to train the CNN model. The controlled experiments conducted using NI 2921 USRPs provided a reliable environment to test the proposed method. The results indicate that the CNN model, trained on our augmented dataset, achieved an accuracy of more than 95\% in classifying the metallic objects, validating the robustness and potential of this approach for industrial applications.
This research highlights the importance of integrating micro-Doppler image analysis with vibration dynamics to achieve precise material classification. The findings suggest that such integration enhances classification accuracy and opens new possibilities for applying machine learning models in detecting and identifying metallic objects across various domains, including defense, space, and security applications. While the current study focuses on the classification of single-layer metallic objects such as brass, copper, and aluminum, the proposed technique has potential applicability to more complex materials, such as multilayer composites. Multilayer composites, which are commonly used in advanced engineering applications, present unique challenges due to their heterogeneous structure and the interaction between layers. These interactions can result in more complex vibration patterns and micro-Doppler signatures, making classification more challenging.

\section*{Acknowledgements}
Salman Liaquat and Nor Muzlifah Mahyuddin would like to thank the Ministry of Higher Education Malaysia for Fundamental Research Grant Scheme with Project Code FRGS/1/2022/ TK07/USM/02/14 for permitting them to carry out this research.
Faran Awais Butt, Ali Hussein Muqaibel and Saleh Alawsh would like to thank King Fahd University of Petroleum and Minerals (KFUPM) for carrying out this research. 

\bibliographystyle{IEEEtran}
\bibliography{References}

\begin{thebibliography}{10}
\providecommand{\url}[1]{#1}
\csname url@samestyle\endcsname
\providecommand{\newblock}{\relax}
\providecommand{\bibinfo}[2]{#2}
\providecommand{\BIBentrySTDinterwordspacing}{\spaceskip=0pt\relax}
\providecommand{\BIBentryALTinterwordstretchfactor}{4}
\providecommand{\BIBentryALTinterwordspacing}{\spaceskip=\fontdimen2\font plus
\BIBentryALTinterwordstretchfactor\fontdimen3\font minus \fontdimen4\font\relax}
\providecommand{\BIBforeignlanguage}[2]{{%
\expandafter\ifx\csname l@#1\endcsname\relax
\typeout{** WARNING: IEEEtran.bst: No hyphenation pattern has been}%
\typeout{** loaded for the language `#1'. Using the pattern for}%
\typeout{** the default language instead.}%
\else
\language=\csname l@#1\endcsname
\fi
#2}}
\providecommand{\BIBdecl}{\relax}
\BIBdecl

\bibitem{huang2021microscopic}
G.~Huang, Q.~Zhang, B.~Zhang, and S.~Li, ``Microscopic mechanism of the combined magnetic-vibration treatment for residual stress reduction,'' \emph{Results in Physics}, vol.~29, p. 104659, 2021.

\bibitem{hanif2022micro}
A.~Hanif, M.~Muaz, A.~Hasan, and M.~Adeel, ``{Micro-Doppler based target recognition with radars: A review},'' \emph{IEEE Sensors Journal}, vol.~22, no.~4, pp. 2948--2961, 2022.

\bibitem{8307105}
Z.~Chen, G.~Li, F.~Fioranelli, and H.~Griffiths, ``{Personnel Recognition and Gait Classification Based on Multistatic Micro-Doppler Signatures Using Deep Convolutional Neural Networks},'' \emph{IEEE Geoscience and Remote Sensing Letters}, vol.~15, no.~5, pp. 669--673, 2018.

\bibitem{muqtadir2021health}
M.~Muqtadir, M.~H. Butt, D.~Qazi, F.~A. Butt, I.~H. Naqvi, and N.~U. Hassan, ``{Health Secure Radar: Use of Micro Doppler Signatures for Health Care and Security Applications},'' in \emph{2021 IEEE VTS 17th Asia Pacific Wireless Communications Symposium (APWCS)}.\hskip 1em plus 0.5em minus 0.4em\relax IEEE, 2021, pp. 1--6.

\bibitem{9004533}
K.~Saho, K.~Inuzuka, and K.~Shioiri, ``{Person Identification Based on Micro-Doppler Signatures of Sit-to-Stand and Stand-to-Sit Movements Using a Convolutional Neural Network},'' \emph{IEEE Sensors Letters}, vol.~4, no.~3, pp. 1--4, 2020.

\bibitem{10004894}
N.~Rojhani, M.~Passafiume, M.~Sadeghibakhi, G.~Collodi, and A.~Cidronali, ``Model-based data augmentation applied to deep learning networks for classification of micro-doppler signatures using fmcw radar,'' \emph{IEEE Transactions on Microwave Theory and Techniques}, vol.~71, no.~5, pp. 2222--2236, 2023.

\bibitem{10401166}
P.~Mandal, L.~P. Roy, and S.~K. Das, ``Flying objects classification based on micro–doppler signature data from uav borne radar,'' \emph{IEEE Geoscience and Remote Sensing Letters}, vol.~21, pp. 1--5, 2024.

\bibitem{khalifa2022comprehensive}
N.~E. Khalifa, M.~Loey, and S.~Mirjalili, ``A comprehensive survey of recent trends in deep learning for digital images augmentation,'' \emph{Artificial Intelligence Review}, pp. 1--27, 2022.

\bibitem{nanni2020data}
L.~Nanni, G.~Maguolo, and M.~Paci, ``Data augmentation approaches for improving animal audio classification,'' \emph{Ecological Informatics}, vol.~57, p. 101084, 2020.

\bibitem{9400955}
S.~Shirmohammadi and H.~Al~Osman, ``Machine learning in measurement part 1: Error contribution and terminology confusion,'' \emph{IEEE Instrumentation \& Measurement Magazine}, vol.~24, no.~2, pp. 84--92, 2021.

\end{thebibliography}

\noindent \textbf{\textit{Salman Liaquat}} (Graduate Student Member, IEEE) (salman.liaquat@student.usm.my) received a Bachelor's degree in Avionics Engineering from the National University of Sciences and Technology (NUST), Risalpur, Pakistan, in 2010 and a Master of Science degree in Avionics Engineering from Air University, Islamabad, Pakistan, in 2019. He is currently pursuing a Ph.D. in Electrical and Electronic Engineering at Universiti Sains Malaysia, Penang. His research interests focus on signal processing for radars and reconfigurable intelligent surfaces.
\vspace{10pt}

\noindent \textbf{\textit{Faran Awais Butt}} (faranawais.butt@kfupm.edu.sa) received the B.Sc. degree in Electrical Engineering from the University of Engineering and Technology (UET) Lahore, Pakistan, in 2009, and the Master’s degree in Computer Engineering from the Lahore University of Management Sciences (LUMS), in 2012. From 2018 to 2019, he worked as a Visiting Researcher at the Radar Group, University College London (UCL), London, U.K. He worked as an Assistant Professor with the University of Management and Technology (UMT), Pakistan from 2019 to 2024 and is currently working as a Post-doctoral fellow with the Interdisciplinary Research Center for Communication Systems and Sensing (IRC-CSS) at King Fahd University of Petroleum and Minerals (KFUPM). His research interests include phased array, MIMO radars, FMCW, Re-configurable Intelligent Surfaces, Phase Codes, and optimization. 
\vspace{10pt}

\noindent \textbf{\textit{Faryal Aurooj Nasir}} (faryalaurooj@gmail.com) received Bachelors degree in Avionics Engineering from the College of Aeronautical engineering, National University of Sciences and Technology Risalpur, Pakistan, in 2010, the master’s degree in Artificial Intelligence from Capital University of Science and Technology, Pakistan, in 2022, and currently doing Ph.D. degree in Artificial Intelligence from Institute of Space Technology, Pakistan. 
\vspace{10pt}

\noindent \textbf{\textit{Ijaz Haider Naqvi}} (Senior Member, IEEE) (ijaznaqvi@lums.edu.pk) received the B.Sc. degree in electrical engineering from the University of Engineering and Technology Lahore, Pakistan, in 2003, the master’s degree in radio communications from SUPELEC Paris, France, in 2006, and the Ph.D. degree in electronics and telecommunications from IETR-INSA Rennes, France, in 2009. He is currently an Associate Professor with the School of Science and Engineering, Lahore University of Management Sciences (LUMS), Pakistan. His current research interests include 5G networks, millimeter wave wireless systems for 5G and 6G telecommunication and multi-antenna radar systems for civilian applications.
\vspace{10pt}

\noindent \textbf{\textit{Nor Muzlifah Mahyuddin}} (Member, IEEE) (eemnmuzlifah@usm.my) received the B.Eng. degree from Universiti Teknologi Malaysia, Malaysia, in 2005, the M.Sc. degree from Universiti Sains Malaysia, Malaysia, in 2006, and the Ph.D. degree from Newcastle University, Newcastle upon Tyne, U.K., in 2011. She is currently an Associate Professor with the School of Electrical and Electronic Engineering, Universiti Sains Malaysia, Malaysia. Her research interests are in the field of the art of miniaturization and low power in RF and microwave engineering, and digital integrated circuits.
\vspace{10pt}

\noindent \textbf{\textit{Ali Hussein Muqaibel}} (Senior Member, IEEE) (muqaibel@kfupm.edu.sa) is a professor in the Electrical Engineering Department at King Fahd University of Petroleum and Minerals (KFUPM). He is the director of the Interdisciplinary Research Center for Communication Systems and Sensing (IRC-CSS). Dr. Muqaibel received his Ph.D. degree from Virginia Polytechnic Institute and State University, Blacksburg, VA, USA, in 2003. He was a Visiting Associate Professor with the Center of Advanced Communications, Villanova University, Villanova, PA, USA, in 2013, a Visiting Professor with the Georgia Institute of Technology in 2015, and a Visiting Scholar with the King Abdullah University for Science and Technology (KAUST), Thuwal, Saudi Arabia, in 2018 and 2019. His research interest spans communications and sensing applications, include direction of arrival estimation, through-wall-imaging, localization, channel characterization, and ultra-wideband signal processing.
\vspace{10pt}

\noindent \textbf{\textit{Saleh Alawsh}} (Member, IEEE) (salawsh@kfupm.edu.sa) received the B.Sc. degree in electronic and communications from Hadhramout University, Al Mukalla, Yemen, in 2007, and the M.Sc. degree in telecommunication and the Ph.D. degree from the King Fahd University of Petroleum and Minerals (KFUPM), Dhahran, Saudi Arabia, in 2013 and 2018, respectively. He is currently a researcher with the Interdisciplinary Research Center for Communication Systems and Sensing (IRC-CSS). His research interests include ultra-wideband systems, narrow band interference mitigation, direction-of-arrival estimation, localization, and compressive sensing.

\end{document}